\documentclass [12,onecolumn]{elsarticle}
\usepackage{graphicx}
\begin{document}
\title{The dependence of the stability of hierarchical triple systems on the orbital
inclination}
\author{Nikolaos Georgakarakos ${^{1}}$}
\ead{georgakarakos@hotmail.com}
\address{${1}$ Higher Technological Educational Institute of Serres, Terma Magnesias, Serres 62124, Greece}

\begin{abstract}
In this paper we study numerically the effect of the initial mutual orbital inclination on
the stability of hierarchical triple systems with initially circular orbits. 
Our aim is to investigate the possibility that the stability boundary may be independent of
the orbital inclination for certain mass ratios. We integrate numerically the 
equations of motion of hierarchical triple systems with initially circular orbits 
and different orbital configurations.  The  mass ratios cover the range from
${10^{-6}}$ to ${10^{6}}$ and the initial mutual inclination angle varies from ${0^{\circ}}$ to
${180^{\circ}}$. The results from the numerical simulations show that for hierarchical triple systems with initially 
circular orbits and for the mass ratios we used, the  initial mutual inclination angle does affect the stability 
boundary.  
\end{abstract}


\maketitle

\section{INTRODUCTION}

The issue of dynamical stability in the context of the three body problem is an intriguing
topic, not just from the purely theoretical point of view, but also because it appears in
many situations in astronomy, such as planetary dynamics or globular cluster evolution.
It is  highly desirable to know whether a triple system is stable or not, but until now, there
has not been any global stability criterion and that is due to a number of reasons.  However, 
throughout the years, various stability criteria have been developed to deal with different 
situations (e.g. see Georgakarakos 2008).    

An effective and useful way of investigating the dynamical stability of a system is by means of
numerical simulations.  The numerical integration of the full equations of motion of a triple 
system has become a popular tool of dynamical astronomy in recent years thanks to the constantly 
improving power of computers.  Besides stellar triple systems or triple systems within the limits of 
our Solar System,  the discovery of more than 850 exoplanets over the past few years has offered us an 
additional field for the use of numerical simulations.  In many cases, the discovered exoplanetary 
systems seem to have different structural characteristics from our Solar System. Planets with eccentric orbits
and 'hot Jupiters' are frequently present in the planetary systems that have been detected so far.  In addition, some 
close-in transiting exoplanets 
have been found to have a significant inclination with respect to the stellar rotation axis (Hebrard et al.
2008, Winn et al. 2009, Triaud et al. 2010, Winn et al 2010). Moreover,  we have the example of the u And
system, where planets c and d have been found to have a mutual inclination of around ${30^{\circ}}$ 
(McArthur et al. 2010).  

The effect of the initial inclination on the stability of triple systems has been part of the discussion 
for some time (e.g. Harrington 1972). Since then, there has been a lot of published work on stability of triple
systems (at an increasing rate in recent years due to the discovery of planetary systems other than ours) 
which would consider mutually inclined orbits.  Some of that work deals with stellar systems, such as for example
Eggleton ${\&}$ Kiseleva (1995), who derived an empirical stability condition for hierarchical stability in triple 
systems, or the more recent work of Valtonen et al. (2008), where the authors tested an analytical stability criterion 
based on the energy change during the encounter between a binary and a single star.  Other work  investigates
the dynamics of planetary systems.  Such examples are the papers of Pilat-Lohinger, Funk ${\&}$ Dvorak (2003) 
and of Doolin ${\&}$ Blundell (2011), where the stability of p-type orbits in 
stellar binaries is investigated, the work of Veras ${\&}$ Armitage (2004) which deals with the dynamics of inclined
two planet systems, the work of Funk et al. (2009), where the authors examine the stability of inclined  orbits of
terrestrial planets in habitable zones, the paper by Marzari, Th${\acute{e}}$bault ${\&}$ Scholl (2009) which deals with planet 
formation in highly inclined stellar binaries, the work of Pilat-Lohinger (2010), which studies the stability of planets in the habitable zone
of inclined multi-planet systems, the paper by Funk et al. (2011), where the authors investigate the long term evolution of inclined 
Earth-like bodies in a system that consists of a star and a gas giant planet.  

In a recent paper (Li, Fu ${\&}$ Sun 2010), it has been suggested that the Hill stability boundary of a hierarchical 
triple system with a low mass inner binary is independent of the initial mutual inclination of the two orbits for certain 
orbital configurations.  They found that when the eccentricity of the outer binary was
not small, the critical semi-major axis ratio was not affected much by the choice of the initial mutual inclination of the hierarchical
triple system and as a result of that, they could obtain a rather simple expression for the critical semi-major axis ratio.  An interesting 
question to answer is whether that holds for general stability as well, i.e. whether
there are certain values of the mass parameters and certain initial orbital configurations for which the
stability boundary is independent or almost independent of the mutual inclination of the two binaries.
Answering that question could be important in many situations in planetary or stellar dynamics, as 
knowing the values of the parameters of a system is often not possible and also, sometimes we  only have 
some statistical knowledge about them. 

Here, we attempt to get an answer to the above question by integrating numerically the equations of motion of
inclined hierarchical triple systems.  We investigate the stability of triple systems over a wide range 
of mass ratios  and at the moment, we restrict ourselves to systems that are on initially circular orbits.

\section{METHOD}

We investigate the stability of hierarchical triple systems by means of numerical simulations. 
By hierarchical we mean a system which consists of a binary (inner binary) and a third body moving on a wider orbit (
the third body and the centre of mass of the inner binary constitute the outer binary).
The systems integrated are both coplanar and non-coplanar, as the aim of this study is to investigate
certain aspects of the effect of the initial orbital inclination on the stability of the system.  We restrict
ourselves to initially circular orbits and the bodies are treated as point masses. No other effects
than Newtonian gravity are taken into consideration. 

We introduce two mass parameters, 
\begin{displaymath}
M_{1}=\frac{m_{2}}{m_{1}+m_{2}} \hspace{1cm}\mbox{and}\hspace{1cm}  M_{2}=\frac{m_{3}}{m_{1}+m_{2}},
\end{displaymath}
where ${m_{1}}$ and ${m_{2}}$ are the masses of the inner binary. For our experiment, 
\begin{displaymath}
M_{1}=0.5,10^{-1},10^{-2},10^{-3},10^{-4},10^{-5},10^{-6} \hspace{0.5cm} \mbox{and}
\end{displaymath}
\begin{displaymath}
M_{2}=10^{-7+i},\hspace{0.2cm} i=1,2,...,13.
\end{displaymath}
Hence,  we have 91 different triple systems in terms of  the masses.  Each system was integrated
for eleven different values of the initial mutual inclination angle, i.e. ${I=0^{\circ}, 20^{\circ}, 40^{\circ},
60^{\circ}, 80^{\circ}, 90^{\circ}, 100^{\circ}, 120^{\circ}, 140^{\circ}, 160^{\circ}, 180^{\circ}}$.  For each
inclination value, the system was started at the following eight positions: ${\phi=0^{\circ},45^{\circ},
90^{\circ},135^{\circ},180^{\circ},225^{\circ},270^{\circ},315^{\circ}}$, where ${\phi}$ is the initial relative 
phase of the two binaries. 

In our experiment, a system was considered to be unstable when we had escape of at least one body, change of hierarchy 
or ejection of one of the bodies during the integration time.  For a given pair of ${M_{1}}$
and ${M_{2}}$ and a specific value of the inclination, we defined four stability categories, depending on
how many of the initial starting positions led to a stable or an unstable configuration. A system fell into the
first category if it was stable at all starting positions. If the system  was stable at five, six or 
seven starting positions, then the system was classified as category two. A system was placed 
in the third category when it was stable at one, two, three or four initial positions and finally, when the system was 
unstable for all starting positions, it was placed in the fourth category.

In order to investigate the stability of a system, we started from a specific value of the initial period ratio 
${X}$ and we progressively moved down to ${X=1}$
by steps of ${0.05}$. We were not just interested to find the first value of ${X}$ for which the system became
unstable, as it is well known that the stability borderline is normally more complicated than that.  In order
to save computation time, we first did some test simulations at high and low inclinations, in order to get an
idea for the choice of the initial period ratio.  Normally, the maximum initial period ratio we decided to start most of
the simulations with was ${X=12}$.  However, in certain cases, we started at lower X, in order to save computation 
time, while in few cases we decided it was necessary to start our simulation with a period ratio ${X>12}$.

The time of integration was chosen such that it would cover the longest period of the system at
least a few times.  For most systems, that time was set to ${10^{4}}$ outer orbit periods.  However, for
some systems with smaller ${M_{2}}$, that integration time was not adequate as the systems had much longer
time of variation of their orbital elements.  Consequently, for those systems, we ran the simulations for
much longer times.  Table 1 gives the time of integration for all our mass ratios.  

\begin{table}
\caption[]{Time of integration for the numerical simulations. The times are given in outer orbital periods and in our system
of units the inner binary period is ${2\pi}$.}
\vspace{0.1 cm}
\begin{center}	
{\small \begin{tabular}{c c c c c c c c c}\hline
${M_{2}}$ &\vline & & & & ${M_{1}}$& & &  \\
 &\vline &0.5&${10^{-1}}$&${10^{-2}}$&${10^{-3}}$&${10^{-4}}$&${10^{-5}}$& ${10^{-6}}$\\
\hline
         
${10^{-6}}$ &  & ${10^{5}}$ & ${10^{5}}$ & ${10^{5}}$ & ${10^{7}}$ & ${10^{7}}$ & ${10^{7}}$ & ${10^{7}}$\\       
${10^{-5}}$ &  & ${10^{5}}$ & ${5\cdot 10^{6}}$ & ${5\cdot 10^{6}}$ & ${5\cdot 10^{6}}$ & ${5\cdot 10^{6}}$ & ${5\cdot 10^{6}}$ & ${5\cdot 10^{6}}$\\     
${10^{-4}}$ &  & ${10^{6}}$ & ${10^{6}}$ & ${10^{6}}$ & ${10^{6}}$ & ${10^{6}}$ & ${10^{6}}$ & ${10^{6}}$\\ 
${10^{-3}}$ &  & ${5\cdot 10^{5}}$ & ${5\cdot 10^{5}}$ & ${5\cdot 10^{5}}$ & ${5\cdot 10^{5}}$ & ${5\cdot 10^{5}}$ & ${5\cdot 10^{5}}$ & ${5\cdot 10^{5}}$\\ 
${10^{-2}}$ &  & ${10^{5}}$ & ${10^{5}}$ & ${10^{5}}$ & ${10^{5}}$ & ${10^{5}}$ & ${10^{5}}$ & ${10^{5}}$\\ 
${10^{-1}-10^{6}}$ &  & ${10^{4}}$ & ${10^{4}}$ & ${10^{4}}$ & ${10^{4}}$ & ${10^{4}}$ & ${10^{4}}$ & ${10^{4}}$\\ 
      
\end{tabular}}
\end{center}	
\end{table}

In order to perform the numerical simulations, we used a symplectic integrator with time 
transformation (Mikkola 1997).  The code uses standard Jacobi coordinates, i.e. it calculates 
the relative position and velocity vectors of the two binaries at every time step.  
Then, by using standard two body formulae, we computed the orbital elements of the two binaries.
The various parameters used by the code, were given the following
values: writing index ${\tt Iwr=1}$, average number of steps per inner binary
period ${\tt NS=25}$, method coefficients ${A_{1}=1}$ and ${A_{2}=15}$,
correction index ${\tt icor=1}$.  We also used units such that
${G=1}$ and ${m_{1}+m_{2}=1}$ and we always started the integrations
with ${a_{1}=1}$, where ${a_{1}}$ is the semi major axis of the inner binary.  We chose the initial 
plane of the inner binary to be our reference plane and the inner binary relative position vector was always started
along the x-axis. Finally, the initial longitude of the ascending node was set equal to zero, as for initially circular orbits
the variation of the initial relative phase is adequate to cover all cases.   In that 
system of units, the initial conditions for the numerical
integrations were as follows:
\begin{displaymath}
r_{1}=1,\hspace{0.5cm} r_{2}=0,\hspace{0.5cm} r_{3}=0,
\end{displaymath}
\begin{displaymath}
R_{1}=a_{2}\cos{\phi},\hspace{0.5cm}R_{2}=a_{2}\sin{\phi}\cos{I},\hspace{0.5cm} R_{3}=a_{2}\sin{\phi}\sin{I}
\end{displaymath}	
\begin{displaymath}
\dot{r}_{1}=0,\hspace{0.5cm} \dot{r}_{2}=1,\hspace{0.5cm} \dot{r}_{3}=0\hspace{0.5cm}
\end{displaymath}
\begin{displaymath}
\dot{R}_{1}=-\sqrt{\frac{M}{a_{2}}}\sin{\phi},\hspace{0.5cm}
\dot{R}_{2}=\sqrt{\frac{M}{a_{2}}}\cos{\phi}\cos{I},\hspace{0.5cm}
\dot{R}_{3}=\sqrt{\frac{M}{a_{2}}}\cos{\phi}\sin{I},
\end{displaymath}	
where ${{\bf{r}}=(r_{1},r_{2},r_{3})}$ and ${{\bf{R}}=(R_{1},R_{2},R_{3})}$ are the relative
position vectors of the inner and outer orbit respectively and ${M=m_{1}+m_{2}+m_{3}}$.

\section{Results}
As we stated in the introduction, our main interest in the simulations we
performed was to see whether there are areas in parameter space where the
stability boundary of a hierarchical triple system is independent of the mutual 
inclination of the two binaries. 

Generally, it is expected that the boundary between
stability and instability will not be very easy to determine, in the sense that
the transition from the one condition to the other could be rather complicated.
That has been noted in many past works (e.g. see Pilat-Lohinger et al. 2003). 
Something similar was seen in our simulations too, as there were situations for which areas of 
stability appeared inside unstable areas and vice versa.  Another thing that was no surprise to us was that 
retrograde orbits, especially those with a mutual inclination close to ${180^{\circ}}$, 
appeared to be more stable than prograde orbits.  Also, it was found that for some systems, the stability
boundary that emerged from the simulations exhibited a similar pattern independently from the choice of the
mass ratios.  More specifically, for ${M_{2} \geq 10}$, the stability limit decreased for ${I=0^{\circ}-40^{\circ}}$,
then it increased for ${I=60^{\circ}-80^{\circ}}$ or ${90^{\circ}}$ and then it deacreased again for the rest of the
mutual inclination angles.

Now, regarding the central issue of our work, which is the dependence of the stability boundary on the 
mutual orbital inclination of the two binaries, for the systems we investigated, the near perpendicular initial
configurations appeared to become unstable at larger initial period ratios compared to the
lower mutual inclination configurations for the same mass ratios.  That is to be expected of 
course, as an even initially circular orbit can become higly eccentric due to the Kozai effect
(Kozai 1962), which can lead to close encounters of the three bodies with possible disruption of the system. 
However, when we started this investigation, we thought that it could be possible for the stability limit 
to be almost unaffected by the variation of the mutual
inclination angle in the case of systems with more extreme mass ratios. After our simulations, that hypothesis seems to
be rejected.  Another possibility is that initially eccentric orbits might result in inclination independent stability 
boundary, but this is not part of the present work.

The above mentioned findings can be seen in Figs.1-4, which  present some results from our simulations for different  mass ratios.  
The plots show the different stability regimes along with the maximum eccentricity values for the smallest initial 
period ratio after which any type of instability appeared for the first time (the given values of the eccentricities are 
the largest ones of the maximum values for each initial relative phase; the eccentricities are rounded to two decimal places).  
More results from our simulations can be found in Table A1 in the appendix.

It is understood that the time of integration is a factor that may have an effect on the results of any numerical simulation. 
Initially, for our experiment, we had chosen an integration time of ${10^{4}}$ outer orbital periods for all our mass ratios.  
That proved to be adequate for most of the systems we investigated, as the longest period of the evolution of each system was covered 
several times by the integration time. The longest period of a given ${M_{1}}$ and ${M_{2}}$ combination varied with the value of the
initial mutual inclination and normally the systems with higher inclination values demonstrated the longest periods.  For example,
for ${M_{2}=0.1}$, ${M_{1}=0.001}$ and ${I=90^{\circ}}$, the integration time of ${10^{4}}$ outer orbit periods was more than two 
hundred times longer than the secular period of the system, while for the same mass ratios and ${I=0^{\circ}}$, the system did not 
show any secular trend.  

For some systems with 
small mass ratios (e.g. ${M_{1}=10^{-6}}$ and ${M_{2}=10^{-6}}$) we found that the stability boundary
exhibited very little change as the intial mutual inclination varied.  However, that proved to be an artefact, as those systems had very
long periods of evolution compared to the ${10^{4}}$ outer orbital periods over which the simulations were performed.  Hence, in those cases, we extended
the time of integration accordingly (basically we had a look at the secular periods of the system for different values of the mutual inclination and 
we chose an integration interval that would be at least a few times longer than the largest secular period we had detected).  The extended results 
for those triple systems showed 
nothing different from what we had found for the rest of the systems, i.e. we found that there was change in the stability boundary as 
the initial mutual inclination varied, with the last stable initial period ratio being larger for higher inclinations.  We also believe
that if the simulations were executed for longer time, the qualitative picture would not alter, but  what would probably happen is that 
the stability boundary for the perpendicular and the near perpendicular configurations would move further outwards. 

The latter hypothesis seems to confirmed by Fig.5 and Fig.6.  In Fig.5, we present a graphical representation of our results for 
${M_{1}=10^{-3}}$ and ${M_{2}=10^{-1}}$.  The plots are similar to the ones of Figs. 1-4, i.e. they show the various stability categories 
for all inclination values and  each inclination angle is 
accompanied by the maximum values of the eccentricities at the stability boundary.  The top plot
is based on simulations that were executed for ${10^{4}}$ outer orbital periods, while the bottom plot shows the results we got for the
same mass ratios but for an integration time of ${10^{5}}$.  Note the change in the stability boundary for the higher mutual inclination
values.  Something similar happens in Fig.6, which is a plot of a system with ${M_{1}=M_{2}=10^{-6}}$. It shows the results of a time extended
numerical integration for two initial mutual inclination angles (${I=0^{\circ}}$ and ${I=90^{\circ}}$). For each inclination angle, the left
column corresponds to an integration time of ${10^{7}}$ outer periods, while the integration time for the right column is ${5\cdot 10^{7}}$ 
outer periods.  It is obvious that the increase of the integration time has only affected the results for the ${90^{\circ}}$.  
Generally, in case of 
performing our simulations over longer time intervals, we would expect to see greater changes in the stability boundary for 
systems with small ${M_{2}}$ and more specifically for those systems of Table 1 which have the longer integration times (${\geq 10^{6}}$), as,
compared to the rest of the systems, they have larger secular period to integration time ratio.

\begin{figure}
\begin{center}
\includegraphics[width=70mm,height=100mm,angle=270]{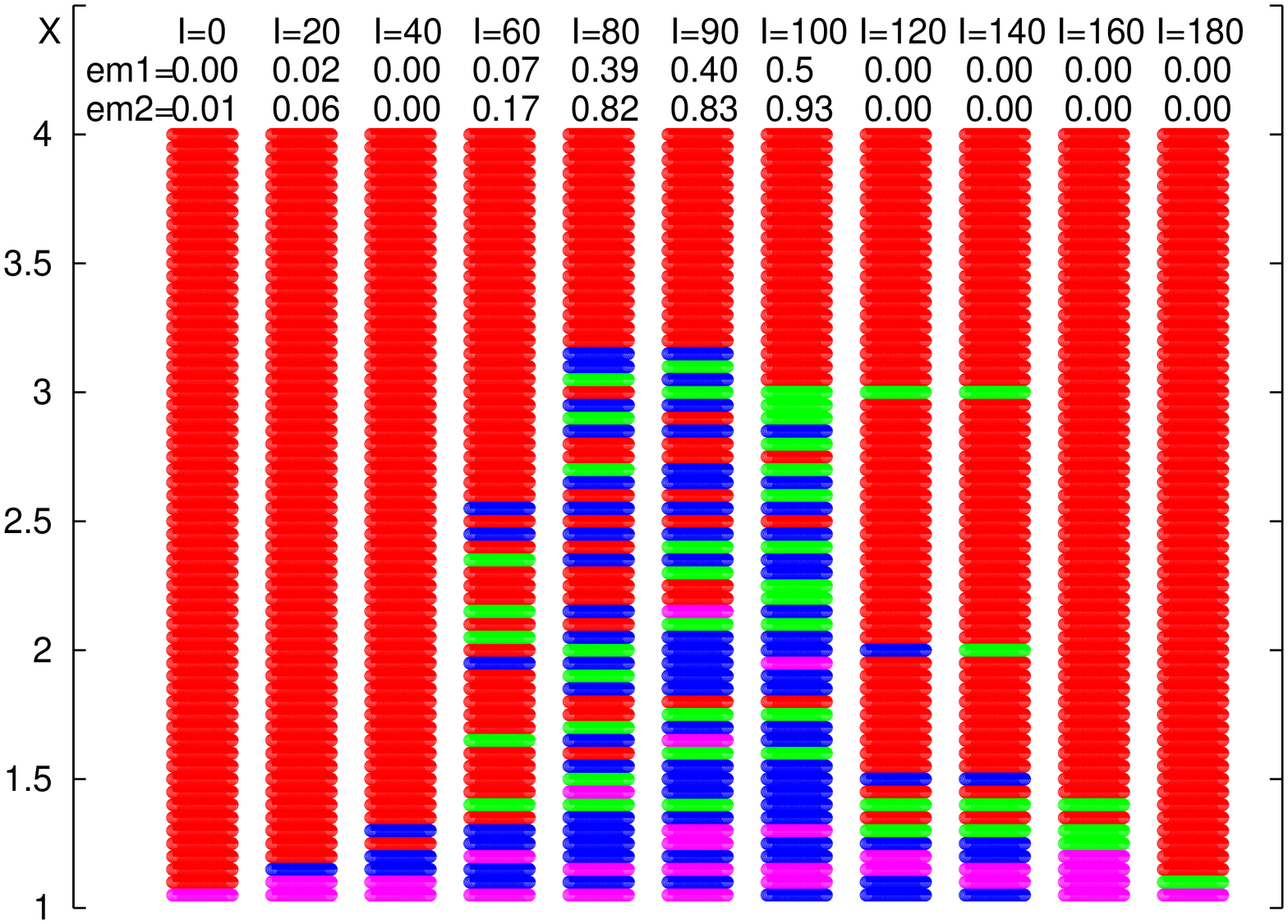}
\includegraphics[width=70mm,height=100mm,angle=270]{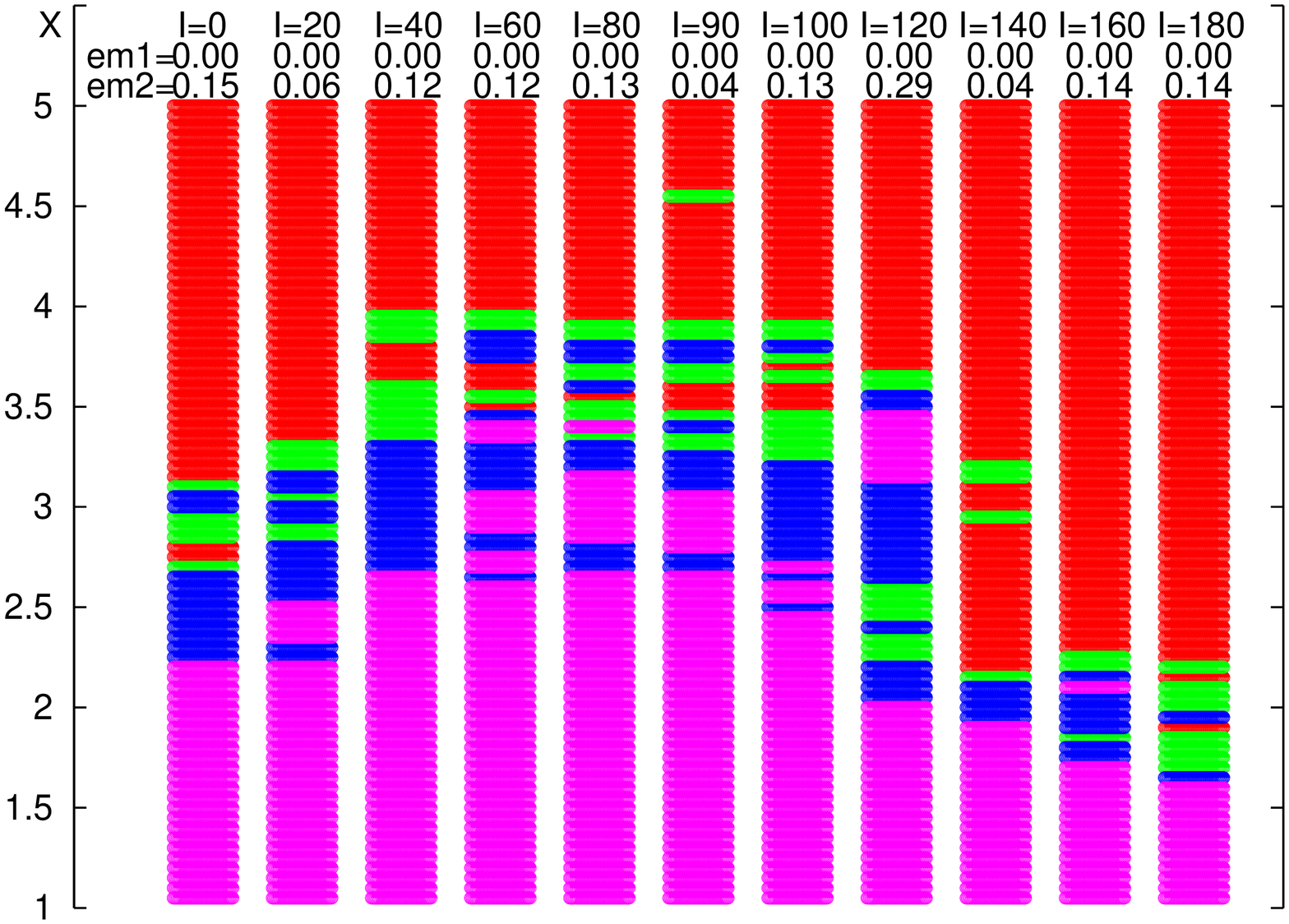}\\
\caption []{Stability-instability graphs for  ${M_{1}=10^{-5}}$, ${M_{2}=10^{-6}}$ (top) and  
${M_{1}=10^{-1}}$, ${M_{2}=10^{-5}}$ (bottom).  The red colour indicates stability 
category one, the green colour indicates stability category two, the blue colour 
indicates stability category three and the purple colour indicates stability 
category four.  Each inclination value is accompanied by the maximum eccentricities 
${em1}$ (inner orbit) and ${em2}$ (outer orbit) at the smallest initial period ratio after which any type of 
instability appeared for the first time.} 
\end{center}
\end{figure}

\begin{figure}
\begin{center}
\includegraphics[width=70mm,height=100mm,angle=270]{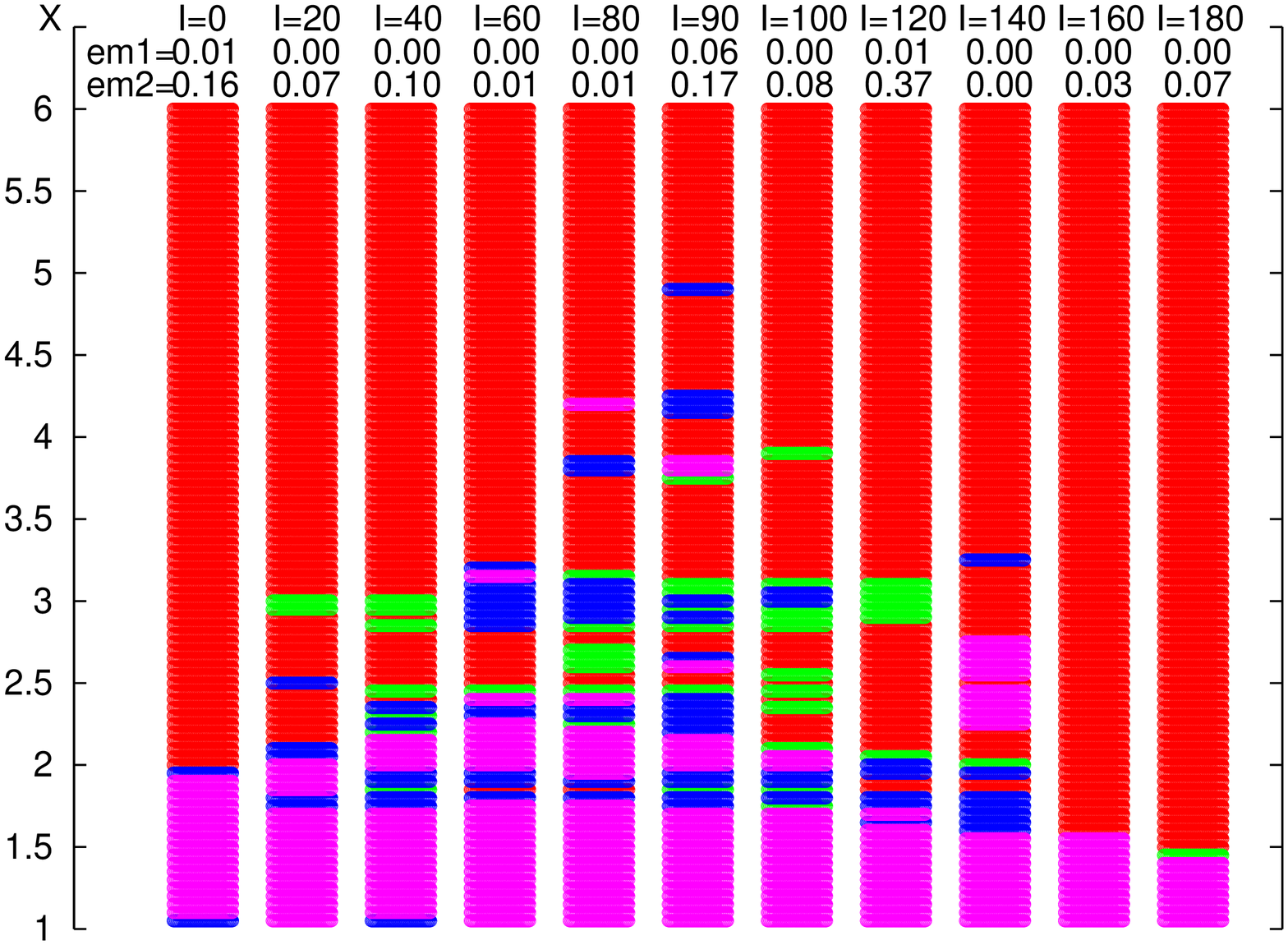}
\includegraphics[width=70mm,height=100mm,angle=270]{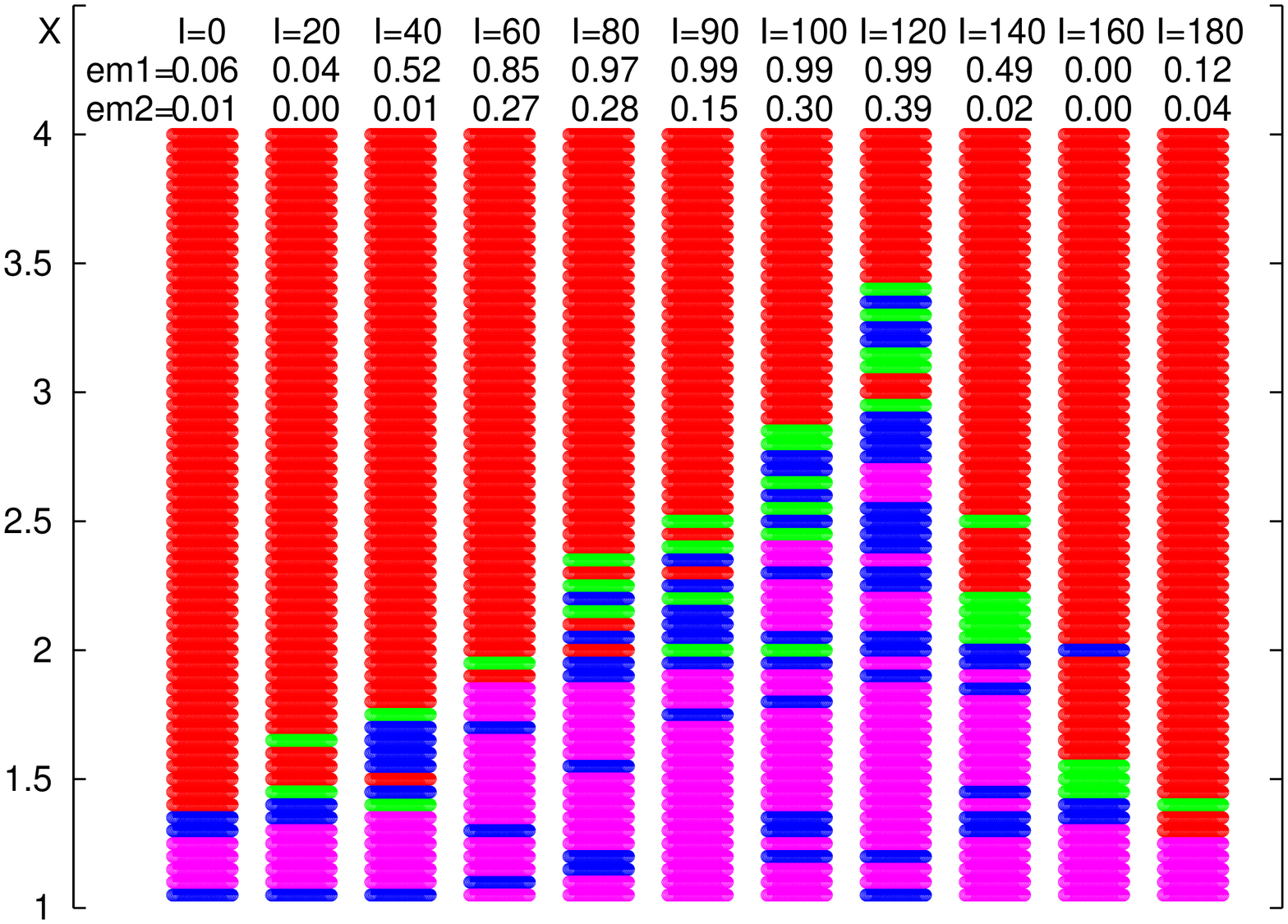}\\
\caption []
{Stability-instability graphs for  ${M_{1}=10^{-2}}$, ${M_{2}=10^{-4}}$ (top) and  
${M_{1}=10^{-4}}$, ${M_{2}=10^{-3}}$ (bottom).  The red colour indicates stability 
category one, the green colour indicates stability category two, the blue colour 
indicates stability category three and the purple colour indicates stability 
category four.  Each inclination value is accompanied by the maximum eccentricities 
${em1}$ (inner orbit) and ${em2}$ (outer orbit) at the smallest initial period ratio after which any type of 
instability appeared for the first time.} 
\end{center}
\end{figure}

\begin{figure}
\begin{center}
\includegraphics[width=70mm,height=100mm,angle=270]{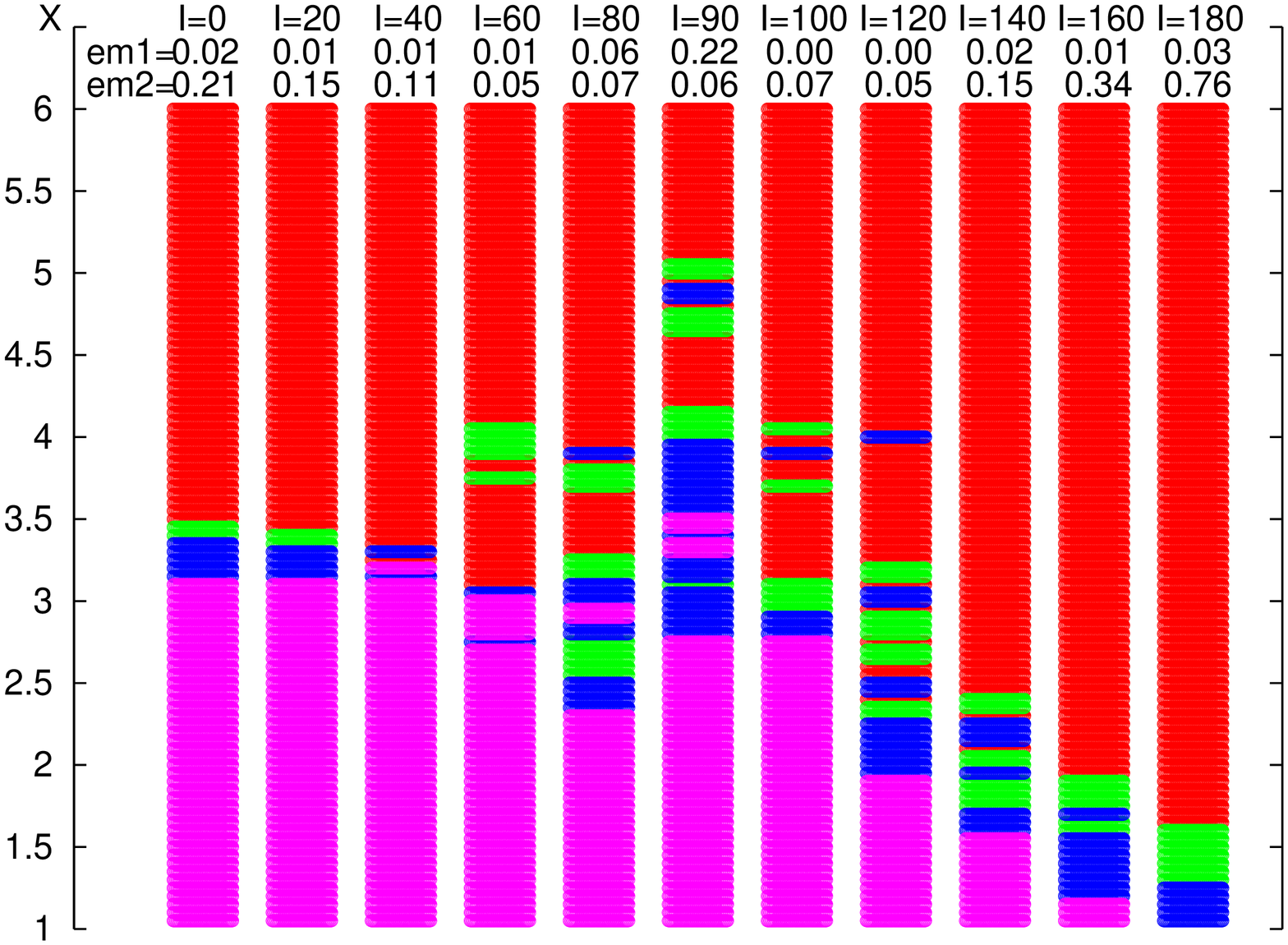}
\includegraphics[width=70mm,height=100mm,angle=270]{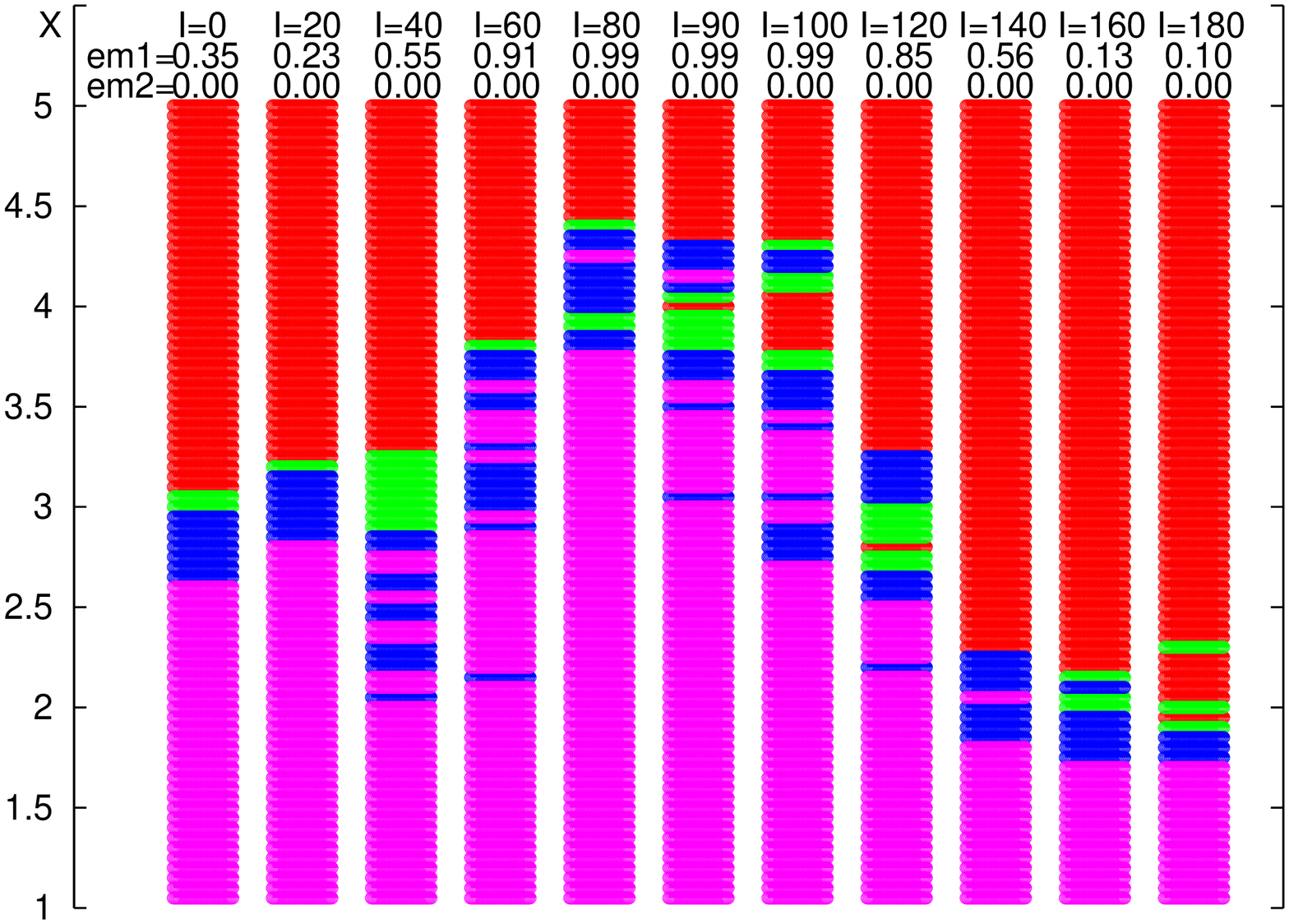}\\
\caption []{Stability-instability graphs for  ${M_{1}=0.5}$, ${M_{2}=10^{-2}}$ (top) and  
${M_{1}=10^{-6}}$, ${M_{2}=10^{-1}}$ (bottom).  The red colour indicates stability 
category one, the green colour indicates stability category two, the blue colour 
indicates stability category three and the purple colour indicates stability 
category four.  Each inclination value is accompanied by the maximum eccentricities 
${em1}$ (inner orbit) and ${em2}$ (outer orbit) at the smallest initial period ratio after which any type of 
instability appeared for the first time.}
\end{center}
\end{figure}

\begin{figure}
\begin{center}
\includegraphics[width=70mm,height=100mm,angle=270]{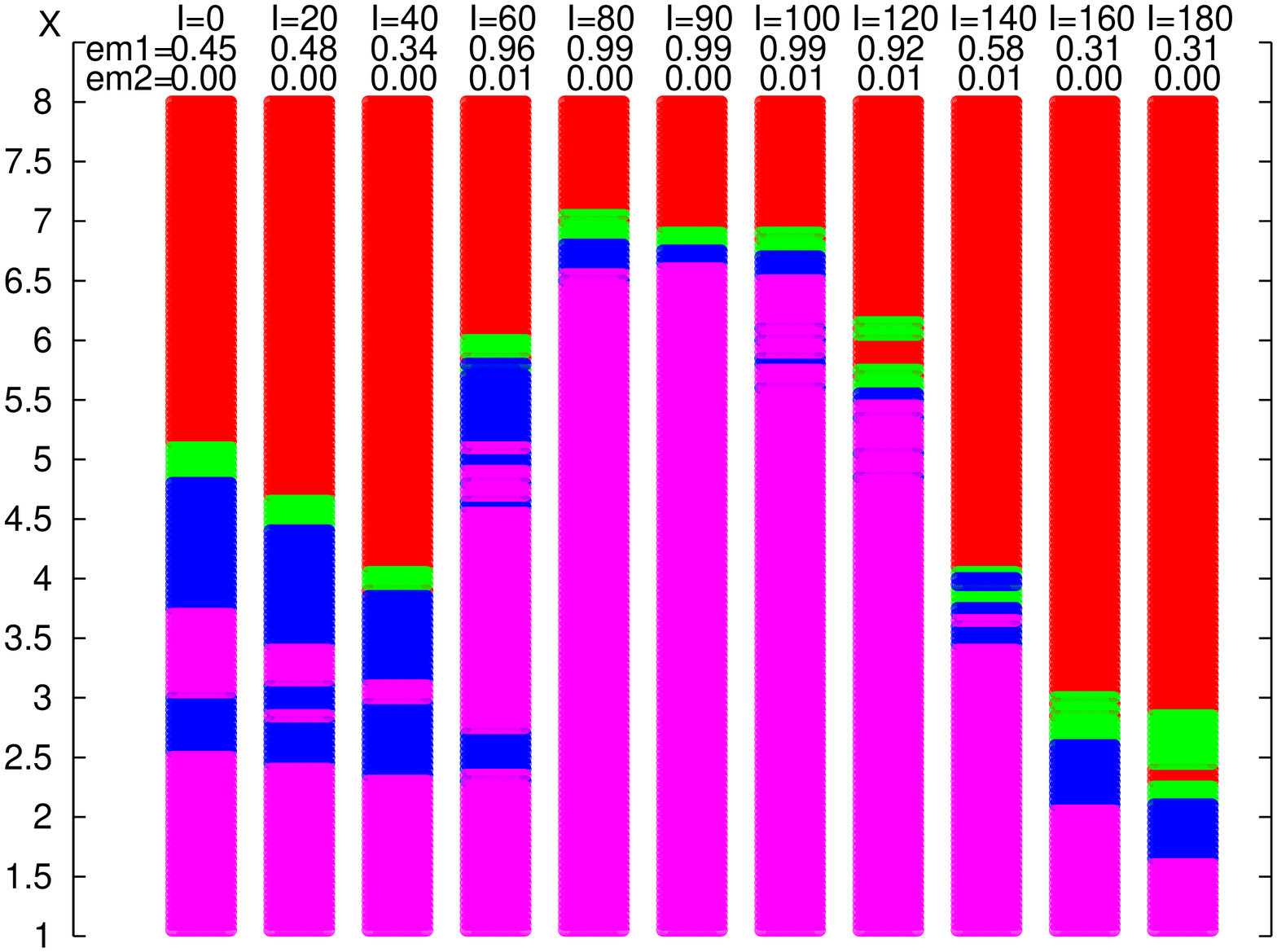}
\includegraphics[width=70mm,height=100mm,angle=270]{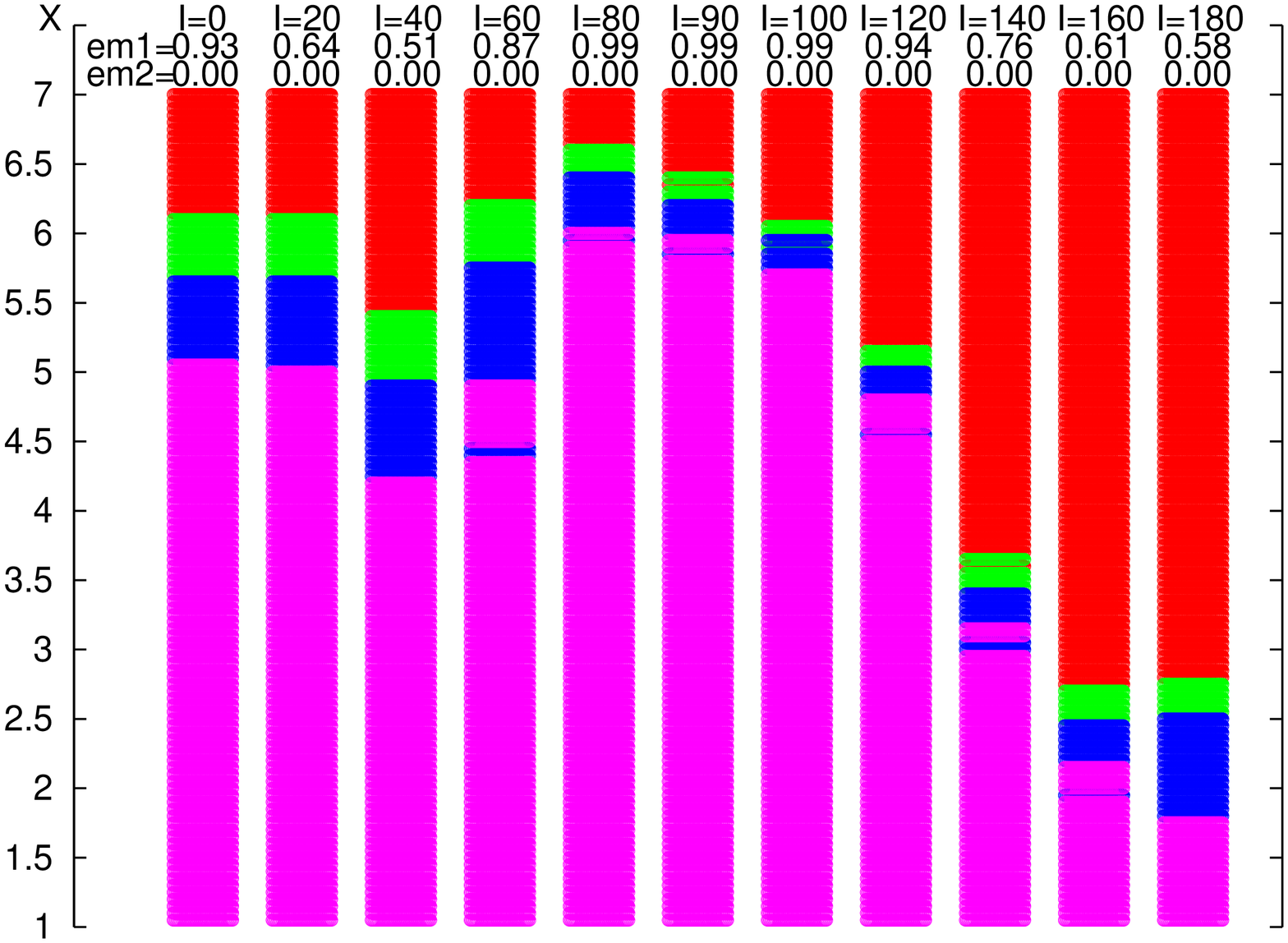}\\
\caption []{Stability-instability graphs for  ${M_{1}=10^{-3}}$, ${M_{2}=1}$ (top) and  
${M_{1}=10^{-2}}$, ${M_{2}=10^{4}}$ (bottom).  The red colour indicates stability 
category one, the green colour indicates stability category two, the blue colour 
indicates stability category three and the purple colour indicates stability 
category four.  Each inclination value is accompanied by the maximum eccentricities 
${em1}$ (inner orbit) and ${em2}$ (outer orbit) at the smallest initial period ratio after which any type of 
instability appeared for the first time.} 
\end{center}
\end{figure}

\begin{figure}
\begin{center}
\includegraphics[width=70mm,height=100mm,angle=270]{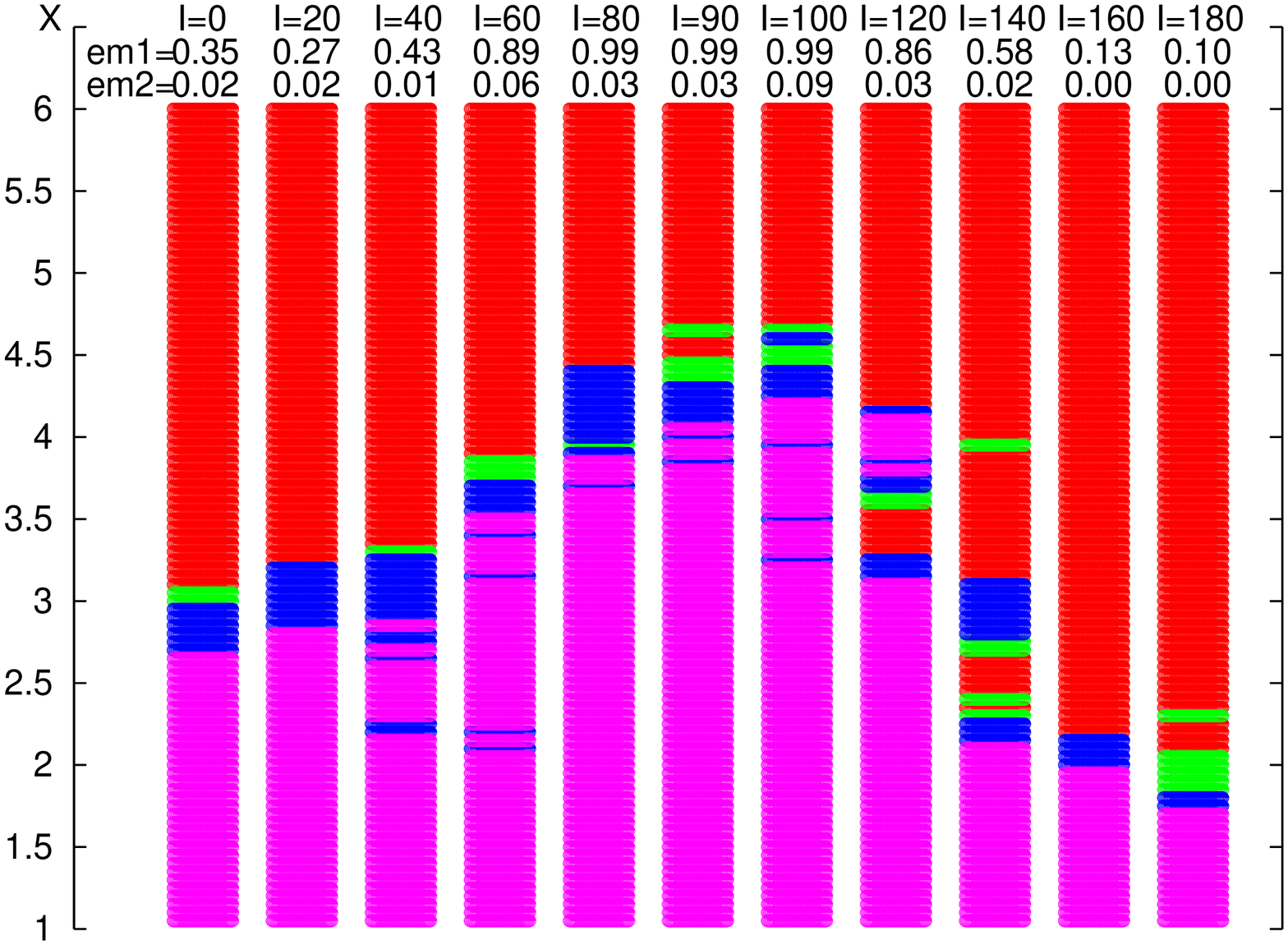}
\includegraphics[width=70mm,height=100mm,angle=270]{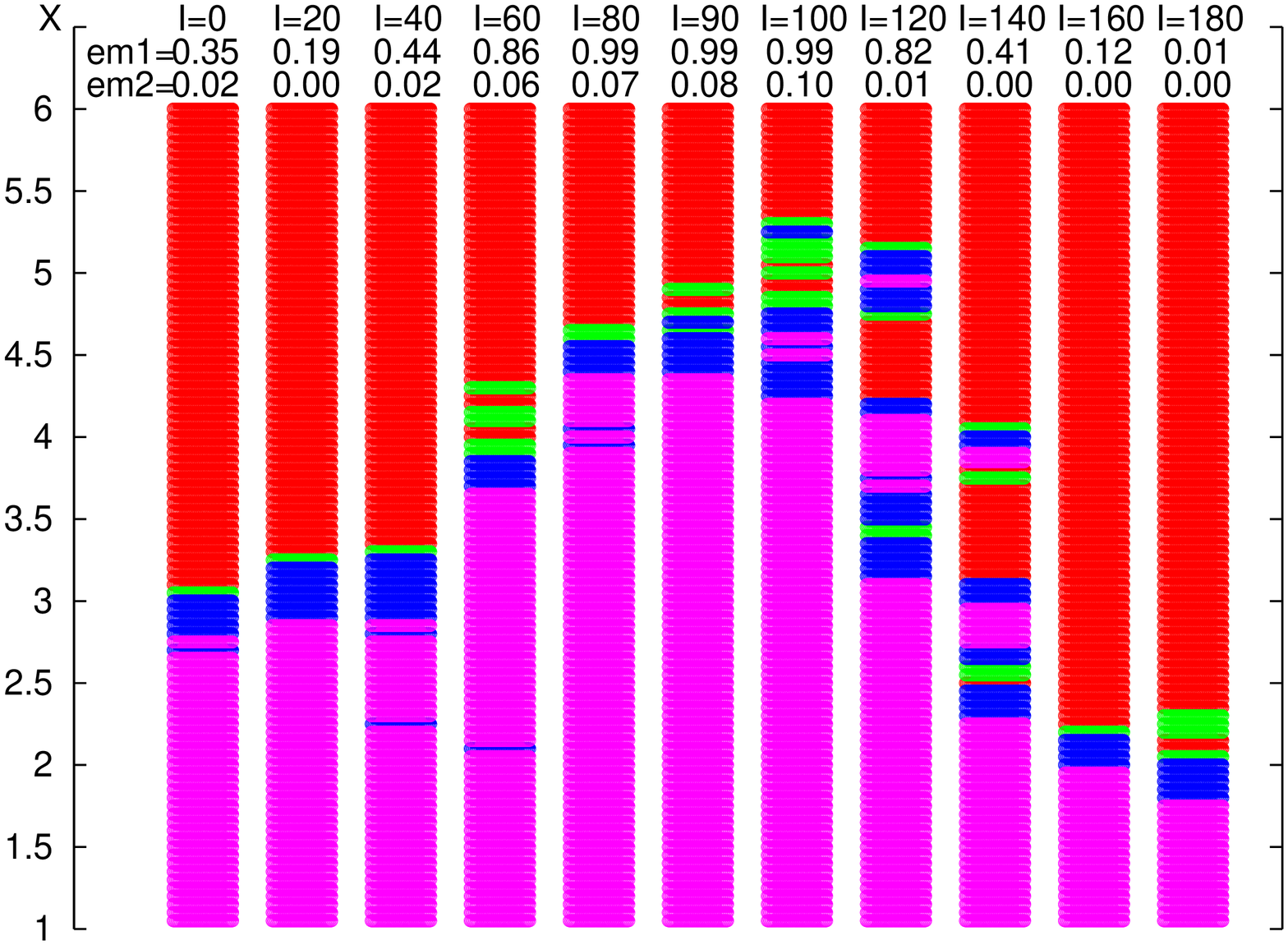}\\
\caption []
{Stability-instability graphs for a system with ${M_{1}=10^{-3}}$ and
${M_{2}=10^{-1}}$. The integration time was ${10^{4}}$ outer orbital periods for the top
graph and ${10^{5}}$ outer orbital periods for the bottom graph.  The red colour 
indicates stability category one, the green colour indicates stability
category two, the blue colour indicates stability category three and the 
purple colour indicates stability category four.  Each inclination value is accompanied
by the maximum eccentricities ${em1}$ (inner orbit) and ${em2}$ (outer orbit) at the the smallest 
initial period ratio 
after which any type of instability appeared for the first time.} 
\end{center}
\end{figure}

\begin{figure}
\begin{center}
\includegraphics[width=70mm,height=100mm,angle=270]{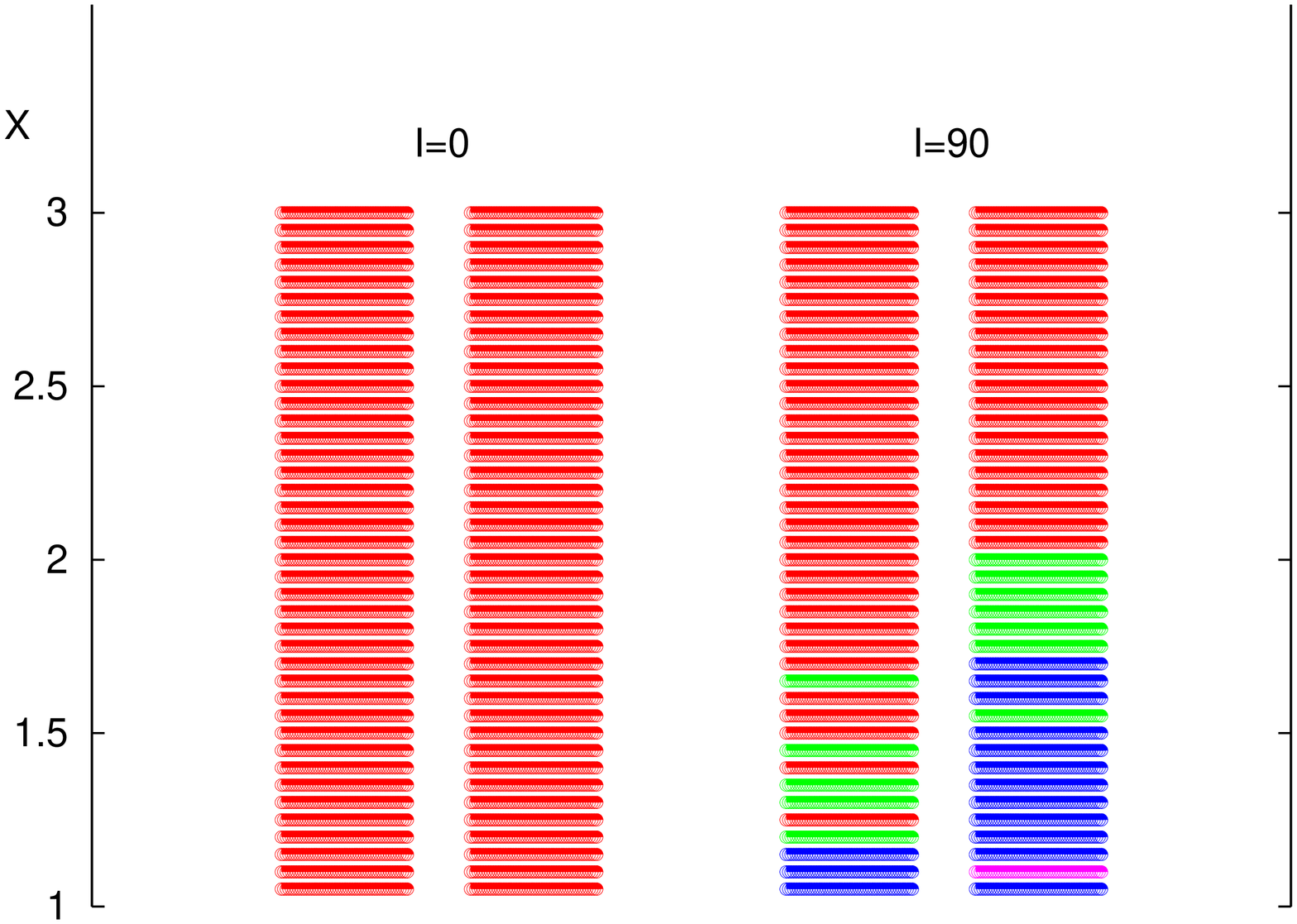}\\
\caption []{Stability-instability graph for a system with ${M_{1}=10^{-6}}$,
${M_{2}=10^{-6}}$, ${I=0^{\circ}}$ and ${I=90^{\circ}}$. The integration time was ${10^{7}}$ 
outer orbital periods for the left column and ${5 \cdot 10^{7}}$ outer 
orbital periods for the right column of each inclination value.  The red colour 
indicates stability category one, the green colour indicates stability
category two, the blue colour indicates stability category three and the 
purple colour indicates stability category four.} 
\end{center}
\end{figure}

\section{Summary}

We have integrated numerically a large number of hierarchical triple systems in
order to investigate their stability behaviour as the  mutual inclination of the
two binary orbits varied. More specifically, we wanted to see whether we could find systems 
for which the stability boundary had little or no dependence at all on the value of the
initial mutual inclination of the two binaries. A recent paper proposed that a situation like that 
is true for the Hill stability of low mass binaries with certain orbital configurations and 
we decided to investigate whether a similar conclusion can be drawn for general stability of a 
hierarchical triple system.
Such a result would be useful in situations where we are interested in the dynamical stability of
a triple system but we do not know the mutual inclination of the two orbits (e.g. an exoplanetary
system).

The results were in agreement with  our previous knowledge, e.g. retrograde orbits seemed
to be more stable, highly inclined systems seemed to be less stable and in many cases there was not a
simple borderline separating stable from unstable motion: islands of stability were found inside unstable areas
and vice versa.   At some point, we thought that for some extreme mass ratio combinations the stability boundary 
was not affected much from the variation of the inclination angle, but that proved to be the result of the choice 
of the integration time.  Therefore, for the mass ratios we investigated and for initially circular orbits, we 
conclude that the stability boundary of a hierarchical triple system does depend on the initial mutual inclination angle of
the two orbits.  The next step of this project is to extend our numerical simulations to systems which have initially 
eccentric orbits.

\section*{ACKNOWLEDGMENTS}
The author would like to thank Seppo Mikkola, who kindly provided the code for integrating HTS and Agamemnon Kehagias
for his assistance regarding some computer hardware issues.

\newpage
\appendix
\section{}

Table A1 presents the critical initial period ratio ${X_{c}}$ for which a system is stable at all initial positions.  
Systems with ${X<X_{c}}$ may demonstrate some form of instability.  Each period ratio is accompanied by an  integer number 
in parentheses which counts the number of times we had a transition
from stability to instability, including the first time. Finally, the difference 
${\delta X_{max}}$ between the maximum and the minimum last stable initial period ratio ${X}$ for all inclinations is also
given in the Table A1.  We would like to say here that ${\delta X_{max}}$
could be affected a bit by the choice of the initial period ratio step we have made,  which was set to 0.05.  A 
smaller step would result in a more accurate picture. However, that does not affect our results qualitatively.

\begin{table}
\caption[]{Critical initial period ratio ${X_{c}}$ for which a system is stable at all initial positions.  
Systems with ${X<X_{c}}$ may demonstrate some form of instability.  For each system, the number in parentheses 
is the number of times we had a transition from stability
to instability. The row after the inclination values gives the difference between the maximum and the minimum last stable 
initial period ratio over all inclinations of a given ${M_{2} - M_{1}}$ pair.}
\begin{center}	
{\small \begin{tabular}{c c c c c c c c c c}\hline
${M_{2}}$&I&\vline& & & &${M_{1}}$& & &  \\
 & &\vline&0.5&${10^{-1}}$&${10^{-2}}$&${10^{-3}}$&${10^{-4}}$&${10^{-5}}$& ${10^{-6}}$\\
\hline
          &${0^{\circ}}$  &  &3.4  (1) &3.15 (3) &2    (1) &1.5  (2) &1.2   (1) &1.1  (1) &1.05 (0) \\
          &${20^{\circ}}$ &  &3.4  (1) &3.35 (2 )&2.55 (2) &1.7  (2) &1.45  (2) &1.2  (1) &1.1  (1)\\
          &${40^{\circ}}$ &  &3.2  (3) &4    (3) &3    (5) &1.85 (3) &1.45  (2) &1.35 (2) &1.15 (1)\\
          &${60^{\circ}}$ &  &3.95 (5) &3.95 (3) &3.2  (3) &3.1  (5) &3.05  (5) &2.6  (9) &1.1  (1)\\
          &${80^{\circ}}$ &  &3.05 (4) &3.95 (2) &3.2  (3) &2.05 (3) &2.05  (4) &3.2 (10) &1.95 (3)\\
${10^{-6}}$&${90^{\circ}}$ &  &2.95 (2) &3.95 (2) &3.15 (5) &2.05 (4) &3.05  (5) &3.2  (7) &1.7  (4)\\
          &${100^{\circ}}$&  &2.65 (1) &3.95 (3) &3.15 (5) &3.05 (5) &3.05  (6) &3.05 (4) &1.75 (2)\\
          &${120^{\circ}}$&  &3.05 (4) &3.65 (2) &3.05 (3) &3.05 (6) &3.05  (6) &3.05 (5) &3.75 (1)\\
          &${140^{\circ}}$&  &2.4  (4) &3    (2) &2.   (3) &3.05 (6) &3.05  (5) &3.05 (5) &2.8  (2)\\
          &${160^{\circ}}$&  &1.95 (1) &2.2  (1) &1.6  (1) &1.55 (2) &1.55  (3) &1.45 (2) &2.05 (2)\\
          &${180^{\circ}}$&  &1.65 (2) &2.15 (2) &1.5  (1) &1.35 (1) &1.3   (2) &1.15 (1) &1.25 (2)\\
          &              &${\delta X_{max}}$=&2.3&1.85&1.7&1.75&1.85&2.15&2.7\\
          &${0^{\circ}}$  &  &3.45  (1) &3.15 (2) &2.05 (1) &1.5  (2) &1.2  (1)  &1.1  (1)  &1.1  (1) \\
          &${20^{\circ}}$ &  &3.4   (1) &3.35 (1) &3.05 (3) &1.8  (3) &1.45 (2)  &1.2  (1)  &1.15 (1)\\
          &${40^{\circ}}$ &  &3.2   (3) &4    (2) &3    (4) &2.3  (5) &1.85 (2)  &1.6  (3)  &1.4 (2)\\
          &${60^{\circ}}$ &  &3.6   (4) &4    (3) &3.25 (3) &3.1  (5) &4.3  (10) &2.35 (7)  &1.6 (4)\\
          &${80^{\circ}}$ &  &3.05  (4) &3.95 (2) &3.9  (3) &3.05 (4) &5.7  (17) &4.1  (17) &3.3 (5)\\
${10^{-5}}$&${90^{\circ}}$ &  &2.95  (2) &4.6  (3) &3.15 (3) &4.1  (7) &6.55 (14) &3    (7)  &1.7 (5)\\
          &${100^{\circ}}$&  &2.75  (2) &3.95 (3) &3.15 (4) &3.05 (5) &6.55 (16) &3.05 (8)  &2.2 (3)\\
          &${120^{\circ}}$&  &3.05  (4) &3.7  (1) &3.4  (4) &3.05 (4) &3.05 (3)  &4.45 (3)  &2.95 (2)\\
          &${140^{\circ}}$&  &2.4   (3) &3.25 (3) &2.75 (5) &3.05 (6) &3.05 (7)  &5.2  (6)  &1.9 (1)\\
          &${160^{\circ}}$&  &1.95  (1) &2.3  (1) &2.05 (3) &1.55 (2) &2.05 (4)  &3.05 (3)  &1.55 (3)\\
          &${180^{\circ}}$&  &1.65  (2) &2.25 (3) &1.5  (1) &1.45 (2) &1.35 (3)  &1.25 (2)  &1.1 (1)\\
          &              &${\delta X_{max}}$=&1.95&2.35&2.4&2.65&5.35&4.1&2.2\\
          &${0^{\circ}}$  &  &3.45 (1) &3.15 (3) &2    (1) &1.5  (2)  &1.25 (1)  &1.2  (1) &1.2  (1) \\
          &${20^{\circ}}$ &  &3.4  (1) &3.35 (2) &3.05 (3) &2.05 (3)  &1.45 (2)  &1.35 (1) &1.25 (1)\\
          &${40^{\circ}}$ &  &3.2  (2) &4    (2) &3.05 (4) &3.05 (5)  &1.85 (4)  &1.45 (2) &1.45 (2)\\
          &${60^{\circ}}$ &  &3.95 (4) &4    (2) &3.25 (3) &4.3  (12) &2.65 (2)  &1.75 (4) &1.6  (3)\\
          &${80^{\circ}}$ &  &3.95 (6) &3.95 (1) &4.25 (6) &7.55 (13) &4.7  (10) &1.75 (2) &1.75 (4)\\
${10^{-4}}$&${90^{\circ}}$ &  &3.75 (4) &4.6  (3) &4.95 (6) &9.1  (17) &4    (7)  &2    (3) &1.75 (4)\\
          &${100^{\circ}}$&  &2.95 (3) &3.95 (3) &3.95 (6) &8.15 (11) &3.2  (6)  &2.35 (2) &1.75 (3)\\
          &${120^{\circ}}$&  &3.25 (5) &3.7  (2) &3.15 (3) &3.1  (5)  &4.1  (3)  &2.95 (1) &1.6  (2)\\
          &${140^{\circ}}$&  &2.45 (2) &3.25 (3) &3.3  (5) &3.05 (9)  &6.15 (9)  &2.   (1) &1.45 (1)\\
          &${160^{\circ}}$&  &1.95 (1) &2.25 (1) &1.6  (1) &2.05 (4)  &3.05 (4)  &1.55 (3) &1.25 (1)\\
          &${180^{\circ}}$&  &1.65 (2) &2.15 (2) &1.5  (1) &1.45 (2)  &1.3  (1)  &1.3  (2) &1.15 (1)\\
          &              &${\delta X_{max}}$=&2.3&2.45&3.45&7.65&4.9&1.75&0.6\\
\end{tabular}}
\end{center}	
\end{table}
\begin{table}
\vspace{0.1 cm}
\begin{center}	
{\small \begin{tabular}{c c c c c c c c c c}
          &${0^{\circ}}$  &  &3.45 (1) &3.15 (2) &2.05 (1)  &1.7  (2)  &1.4  (1) &1.4  (1) &1.4  (1) \\
          &${20^{\circ}}$ &  &3.4  (1) &3.35 (1) &3.1  (3)  &2    (3)  &1.7  (2) &1.5  (1) &1.5  (1)\\
          &${40^{\circ}}$ &  &3.25 (1) &5.05 (3) &5.05 (5)  &3.05 (3)  &1.8  (2) &1.8  (2) &1.8  (2)\\
          &${60^{\circ}}$ &  &4.05 (5) &3.95 (2) &3.8  (2)  &5.05 (7)  &2    (2) &1.75 (1) &1.85 (2)\\
          &${80^{\circ}}$ &  &3.95 (6) &3.95 (1) &10.1 (11) &8    (8)  &2.4  (4) &2    (2) &1.9  (2)\\
${10^{-3}}$&${90^{\circ}}$ &  &4.1  (6) &4.85 (2) &12.1 (10) &5.15 (5)  &2.55 (3) &2    (2) &2    (2)\\
          &${100^{\circ}}$&  &2.95 (2) &4.4  (4) &10.1 (7)  &5.1  (3)  &2.9  (1) &2.1  (2) &2    (2)\\
          &${120^{\circ}}$&  &3.25 (5) &3.75 (1) &3.25 (3)  &5.1  (3)  &3.45 (2) &2.05 (2) &1.85 (1)\\
          &${140^{\circ}}$&  &2.45 (2) &5.15 (7) &3.1  (5)  &7.8  (18) &2.55 (2) &1.65 (1) &1.6  (1)\\
          &${160^{\circ}}$&  &1.95 (1) &3.1  (2) &2.05 (2)  &3.05 (3)  &2.05 (2) &1.55 (1) &1.5  (1)\\
          &${180^{\circ}}$&  &1.65 (1) &2.15 (2) &1.6  (2)  &1.5  (2)  &1.45 (2) &1.35 (1) &1.3  (1)\\
          &              &${\delta X_{max}}$=&2.45&3&10.05&6.5&2.05&0.75&0.7\\
          &${0^{\circ}}$  &  &3.5  (1) &3.2   (1) &2.1   (1) &1.95 (1) &1.95  (1) &1.95 (1) &1.9  (1) \\
          &${20^{\circ}}$ &  &3.45 (1) &4.2   (2) &3.15  (3) &2.05 (1) &1.9   (1) &1.95 (1) &1.95 (1)\\
          &${40^{\circ}}$ &  &3.35 (2) &5.15  (3) &4     (5) &2.05 (1) &2     (1) &1.95 (1) &1.95 (1)\\
          &${60^{\circ}}$ &  &4.1  (3) &3.95  (1) &7.05  (6) &3.2  (2) &2.6   (3) &2.6  (3) &2.6  (3)\\
          &${80^{\circ}}$ &  &3.95 (3) &11    (5) &10.05 (5) &3.2  (2) &2.6   (3) &2.3  (1) &2.25 (1)\\
${10^{-2}}$&${90^{\circ}}$ &  &5.1  (4) &11.05 (4) &7.1   (4) &4. 1 (2) &2.6   (2) &2.5  (2) &2.25 (1)\\
          &${100^{\circ}}$&  &4.1  (4) &12.05 (4) &6.1   (3) &4.15 (4) &2.65  (1) &2.25 (2) &2.25 (1)\\
          &${120^{\circ}}$&  &4.05 (7) &3.95  (1) &5.25  (2) &5.05 (3) &2.55  (2) &2.15 (2) &2.15 (2)\\
          &${140^{\circ}}$&  &2.45 (3) &5.1   (6) &7.75  (4) &3.05 (3) &2.1   (2) &1.9  (1) &1.95 (2)\\
          &${160^{\circ}}$&  &1.95 (1) &3.2   (2) &3.05  (3) &2.05 (2) &2.05  (2) &2.05 (2) &1.65 (1)\\
          &${180^{\circ}}$&  &1.65 (1) &2.1   (4) &1.6   (1) &1.55 (1) &1.55  (1) &1.55 (1) &1.55 (1)\\
          &              &${\delta X_{max}}$=&3.45&9.95&8.45&3.5&1.1&1.05&1.05\\
          &${0^{\circ}}$  &  &4.4  (3) &4.15 (3) &3.2  (1) &3.1  (1) &3.05 (1) &3.1  (1) &3.1  (1) \\
          &${20^{\circ}}$ &  &4.4  (4) &4.3  (2) &3.3  (1) &3.25 (1) &3.25 (1) &3.25 (1) &3.25 (1)\\
          &${40^{\circ}}$ &  &4.25 (2) &5.2  (2) &3.95 (2) &3.35 (1) &3.35 (1) &3.35 (1) &3.3  (1)\\
          &${60^{\circ}}$ &  &6.   (4) &7.15 (4) &4.35 (1) &3.9  (1) &3.9  (1) &3.9  (1) &3.85 (1)\\
          &${80^{\circ}}$ &  &6.1  (2) &10.1 (3) &5.3  (1) &4.45 (1) &4.45 (1) &4.4  (1) &4.45 (1)\\
${10^{-1}}$&${90^{\circ}}$ &  &6.15 (2) &8.05 (4) &6.25 (4) &4.7  (2) &4.35 (1) &4.45 (2) &4.35 (2)\\
          &${100^{\circ}}$&  &6.15 (2) &7.95 (3) &6.95 (3) &4.7  (1) &4.35 (1) &4.3  (2) &4.35 (2)\\
          &${120^{\circ}}$&  &6.9  (3) &7    (4) &6.1  (3) &4.2  (2) &4.2  (4) &4.2  (2) &3.3  (2)\\
          &${140^{\circ}}$&  &6.2  (2) &5.8  (2) &4.1  (2) &4    (4) &3.15 (2) &2.3  (1) &2.3  (1)\\
          &${160^{\circ}}$&  &6.05 (3) &3.15 (2) &2.25 (1) &2.2  (1) &2.2  (1) &2.2  (1) &2.2  (1)\\
          &${180^{\circ}}$&  &1.6  (2) &1.85 (1) &2.15 (2) &2.35 (2) &2.35 (3) &2.1  (2) &2.35 (3)\\
          &              &${\delta X_{max}}$=&5.3&8.25&4.8&2.5&2.25&2.35&2.25\\
          &${0^{\circ}}$  &  &4.45 (1) &5    (1) &5.15 (1) &5.15 (1) &5.15 (1) &5.15 (1) &5.15 (1)\\
          &${20^{\circ}}$ &  &4.35 (1) &4.85 (2) &4.7  (1) &4.7  (1) &4.7  (1) &4.7  (1) &4.7  (1)\\
          &${40^{\circ}}$ &  &4.2  (1) &4.6  (2) &4.35 (2) &4.1  (2) &4.1  (2) &4.1  (2) &4.1  (2)\\
          &${60^{\circ}}$ &  &5.4  (2) &6.15 (2) &6    (1) &6.05 (2) &6.15 (2) &6.05 (1) &6    (1)\\
          &${80^{\circ}}$ &  &6.4  (1) &7.65 (3) &7    (1) &7.1  (2) &6.95 (1) &7    (1) &7    (1)\\
${1}$     &${90^{\circ}}$ &  &6.3  (1) &8.3  (3) &7.5  (5) &8.7  (2) &6.9  (1) &7    (1) &6.95 (1)\\
          &${100^{\circ}}$&  &6.   (4) &9.7  (1) &7.75 (5) &6.95 (2) &6.85 (1) &6.9  (2) &6.85 (2)\\
          &${120^{\circ}}$&  &5.1  (1) &6.5  (2) &6.3  (3) &6.2  (4) &5.85 (1) &5.8  (2) &5.75 (2)\\
          &${140^{\circ}}$&  &3.9  (3) &4.05 (1) &4.1  (1) &4.1  (1) &4.1  (1) &4.1  (1) &4.1  (1)\\
          &${160^{\circ}}$&  &2.7  (2) &2.95 (3) &3.05 (3) &3.05 (3) &3.05 (3) &3.05 (2) &3.05 (3)\\
          &${180^{\circ}}$&  &2.85 (6) &2.9  (2) &2.9  (2) &2.9  (2) &2.9  (2) &2.9  (2) &2.9  (2)\\
          &              &${\delta X_{max}}$=&3.7&6.8&4.85&5.8&4.05&4.1&4.1\\
\end{tabular}}
\end{center}	
\end{table}
\begin{table}
\vspace{0.1 cm}
\begin{center}	
{\small \begin{tabular}{c c c c c c c c c c}
          &${0^{\circ}}$  &  &6.1  (1) &6.6  (1) &6.7  (1) &6.7  (1) &6.7  (1) &6.7  (1) &6.7  (1) \\
          &${20^{\circ}}$ &  &5.85 (1) &6.25 (1) &6.35 (1) &6.35 (1) &6.35 (1) &6.4  (1) &6.35 (1)\\
          &${40^{\circ}}$ &  &5.55 (2) &5.4  (1) &5.4  (1) &5.4  (1) &5.4  (1) &5.4  (1) &5.4  (1)\\
          &${60^{\circ}}$ &  &6.45 (2) &6.8  (3) &6.7  (1) &6.95 (3) &6.75 (1) &6.75 (1) &6.8  (1)\\
          &${80^{\circ}}$ &  &6.45 (2) &7.35 (2) &7.45 (1) &7.6  (2) &7.65 (2) &7.5  (1) &7.6  (1)\\
${10}$    &${90^{\circ}}$ &  &6.3  (2) &7.35 (2) &7.55 (1) &7.7  (3) &7.6  (2) &7.45 (1) &7.65 (3)\\
          &${100^{\circ}}$&  &5.9  (1) &7.25 (3) &8    (3) &7.2  (1) &7.25 (1) &7.3  (2) &7.45 (2)\\
          &${120^{\circ}}$&  &5.1  (1) &6.6  (1) &6.4  (3) &6.3  (2) &6.3  (3) &6.25 (1) &6.2  (1)\\
          &${140^{\circ}}$&  &3.55 (1) &4.3  (2) &4.35 (1) &4.35 (2) &4.25 (1) &4.45 (2) &4.3  (1)\\
          &${160^{\circ}}$&  &2.75 (1) &3.15 (2) &3.1  (2) &3.1  (2) &3.1  (2) &3.1  (2) &3.1  (2)\\
          &${180^{\circ}}$&  &2.95 (1) &3.25 (1) &3.   (1) &2.95 (1) &2.95 (1) &2.95 (1) &2.95 (1)\\
          &              &${\delta X_{max}}$=&3.7&4.2&5&4.75&4.7&4.55&4.7\\
          &${0^{\circ}}$  &  &6.1  (1) &6.45 (1) &6.5  (1) &6.5  (1) &6.5  (1) &6.5  (1) &6.5  (1) \\
          &${20^{\circ}}$ &  &6.1  (1) &6.35 (1) &6.35 (1) &6.4  (1) &6.35 (1) &6.4  (1) &6.4  (1)\\
          &${40^{\circ}}$ &  &5.5  (2) &5.7  (2) &5.55 (1) &5.55 (1) &5.55 (1) &5.55 (1) &5.55 (1)\\
          &${60^{\circ}}$ &  &6.25 (1) &6.55 (2) &6.5  (2) &6.6  (2) &6.55 (1) &6.6  (2) &6.5  (1)\\
          &${80^{\circ}}$ &  &6.5  (1) &7.05 (1) &7.15 (1) &7.15 (2) &7.15 (1) &7.05 (1) &7.05 (1)\\
${10^{2}}$ &${90^{\circ}}$ &  &6.3  (2) &6.9  (1) &7.1  (2) &7.15 (2) &7.15 (4) &7.1  (2) &7.05 (2)\\
          &${100^{\circ}}$&  &5.95 (1) &6.65 (2) &6.8  (2) &6.9  (2) &6.75 (2) &6.7  (1) &6.8  (2)\\
          &${120^{\circ}}$&  &5.15 (2) &5.6  (1) &5.7  (1) &5.7  (1) &5.65 (1) &5.7  (1) &5.85 (2)\\
          &${140^{\circ}}$&  &3.45 (1) &3.9  (1) &4.05 (2) &3.95 (2) &4    (2) &4    (2) &3.95 (1)\\
          &${160^{\circ}}$&  &2.7  (1) &3.05 (2) &3.05 (2) &3.05 (2) &3.05 (2) &3.05 (2) &3.05 (2)\\
          &${180^{\circ}}$&  &2.8  (1) &2.85 (1) &2.85 (1) &2.85 (1) &2.85 (1) &2.85 (1) &2.85 (1)\\
          &              &${\delta X_{max}}$=&3.8&4.2&4.3&4.3&4.3&4.25&4.2\\
          &${0^{\circ}}$  &  &6.1  (1) &6.25 (1) &6.25 (1) &6.3  (1) &6.3  (1) &6.3  (1) &6.3  (1)\\
          &${20^{\circ}}$ &  &6.05 (1) &6.2  (1) &6.2  (1) &6.25 (1) &6.25 (1) &6.25 (1) &6.25 (1)\\
          &${40^{\circ}}$ &  &5.5  (1) &5.55 (1) &5.55 (1) &5.5  (1) &5.55 (2) &5.55 (1) &5.55 (1)\\
          &${60^{\circ}}$ &  &6.15 (1) &6.35 (2) &6.45 (2) &6.35 (1) &6.45 (2) &6.45 (2) &6.45 (2)\\
          &${80^{\circ}}$ &  &6.55 (2) &6.8  (3) &6.85 (1) &6.85 (1) &6.75 (1) &6.75 (1) &6.8  (1)\\
${10^{3}}$ &${90^{\circ}}$ &  &6.25 (1) &6.5  (1) &6.55 (1) &6.55 (1) &6.6  (1) &6.6  (2) &6.55 (1)\\
          &${100^{\circ}}$&  &6    (1) &6.2  (1) &6.35 (1) &6.35 (2) &6.4  (2) &6.25 (1) &6.25 (1)\\
          &${120^{\circ}}$&  &5.05 (1) &5.3  (1) &5.35 (1) &5.4  (1) &5.35 (1) &5.4  (2) &5.45 (2)\\
          &${140^{\circ}}$&  &3.45 (1) &3.75 (1) &3.75 (1) &3.75 (1) &3.75 (1) &3.75 (1) &3.85 (2)\\
          &${160^{\circ}}$&  &2.7  (1) &2.7  (1) &2.75 (1) &2.75 (1) &2.75 (1) &2.75 (1) &2.75 (1)\\
          &${180^{\circ}}$&  &2.8  (1) &2.8  (1) &2.8  (1) &2.8  (1) &2.8  (1) &2.8  (1) &2.8  (1)\\
          &              &${\delta X_{max}}$=&3.85&4.1&4.1&4.1&4&4&4.05\\
          &${0^{\circ}}$  &  &6.05 (1) &6.15 (1) &6.15 (1) &6.15 (1) &6.15 (1) &6.15 (1) &6.15 (1)\\
          &${20^{\circ}}$ &  &6.05 (1) &6.1  (1) &6.15 (1) &6.15 (1) &6.15 (1) &6.15 (1) &6.2  (1)\\
          &${40^{\circ}}$ &  &5.4  (1) &5.45 (1) &5.45 (1) &5.45 (1) &5.45 (1) &5.45 (1) &5.45 (1)\\
          &${60^{\circ}}$ &  &6.25 (1) &6.25 (1) &6.25 (1) &6.25 (1) &6.3  (1) &6.25 (1) &6.25 (1)\\
          &${80^{\circ}}$ &  &6.45 (1) &6.55 (1) &6.65 (1) &6.7  (2) &6.7  (2) &6.7  (2) &6.55 (1)\\
${10^{4}}$ &${90^{\circ}}$ &  &6.25 (1) &6.45 (2) &6.45 (2) &6.5  (2) &6.45 (2) &6.5  (2) &6.45 (2)\\
          &${100^{\circ}}$&  &5.95 (1) &6.15 (2) &6.1  (1) &6.15 (1) &6.15 (2) &6.1  (1) &6.05 (1)\\
          &${120^{\circ}}$&  &5.   (1) &5.2  (1) &5.2  (1) &5.35 (2) &5.25 (2) &5.25 (1) &5.2  (1)\\
          &${140^{\circ}}$&  &3.45 (1) &3.75 (2) &3.7  (2) &3.6  (1) &3.6  (1) &3.7  (2) &3.7  (2)\\
          &${160^{\circ}}$&  &2.7  (1) &2.7  (1) &2.75 (1) &2.75 (1) &2.75 (1) &2.75 (1) &2.75 (1)\\
          &${180^{\circ}}$&  &2.8  (1) &2.8  (1) &2.8  (1) &2.8  (1) &2.8  (1) &2.8  (1) &2.8  (1)\\
          &              &${\delta X_{max}}$=&3.75&3.85&3.9&3.95&3.95&3.95&3.8\\
\end{tabular}}
\end{center}	
\end{table}
\begin{table}
\vspace{0.1 cm}
\begin{center}	
{\small \begin{tabular}{c c c c c c c c c c}
          &${0^{\circ}}$  &  &6.05 (1) &6.1  (1) &6.1  (1) &6.1  (1) &6.1  (1) &6.1  (1) &6.1  (1) \\
          &${20^{\circ}}$ &  &6.05 (1) &6.1  (1) &6.1  (1) &6.1  (1) &6.1  (1) &6.1  (1) &6.1  (1)\\
          &${40^{\circ}}$ &  &5.45 (1) &5.45 (1) &5.45 (1) &5.45 (1) &5.45 (1) &5.45 (1) &5.45 (1)\\
          &${60^{\circ}}$ &  &6.2  (1) &6.25 (1) &6.2  (1) &6.2  (1) &6.2  (1) &6.25 (1) &6.2  (1)\\
          &${80^{\circ}}$ &  &6.45 (1) &6.55 (1) &6.5  (1) &6.5  (1) &6.6  (2) &6.45 (1) &6.5  (1)\\
${10^{5}}$ &${90^{\circ}}$ &  &6.3  (1) &6.35 (2) &6.35 (1) &6.3  (1) &6.4  (2) &6.3  (2) &6.35 (1)\\
          &${100^{\circ}}$&  &6    (2) &6.05 (1) &6.   (1) &6    (1) &6    (1) &6.05 (1) &6.05 (2)\\
          &${120^{\circ}}$&  &5.1  (1) &5.1  (1) &5.1  (1) &5.2  (2) &5.15 (1) &5.15 (1) &5.15 (1)\\
          &${140^{\circ}}$&  &3.45 (1) &3.5  (1) &3.5  (1) &3.7  (2) &3.55 (1) &3.7  (2) &3.55 (1)\\
          &${160^{\circ}}$&  &2.65 (1) &2.65 (1) &2.75 (1) &2.7  (1) &2.7  (1) &2.7  (1) &2.7  (1)\\
          &${180^{\circ}}$&  &2.8  (1) &2.8  (1) &2.8  (1) &2.8  (1) &2.8  (1) &2.8  (1) &2.8  (1)\\
          &              &${\delta X_{max}}$=&3.8&3.9&3.75&3.8&3.9&3.75&3.8\\
          &${0^{\circ}}$  &  &6.05 (1) &6.1  (1) &6.1  (1) &6.1  (1) &6.1  (1) &6.1  (1)  &6.1  (1) \\
          &${20^{\circ}}$ &  &6.05 (1) &6.05 (1) &6.1  (1) &6.1  (1) &6.1  (1) &6.1  (1)  &6.1  (1)\\
          &${40^{\circ}}$ &  &5.45 (1) &5.4  (1) &5.45 (1) &5.45 (1) &5.4  (1) &5.45 (1)  &5.45 (1)\\
          &${60^{\circ}}$ &  &6.2  (1) &6.2  (1) &6.2  (1) &6.2  (1) &6.25 (1) &6.2  (1)  &6.2  (1)\\
          &${80^{\circ}}$ &  &6.45 (1) &6.5  (1) &6.5  (1) &6.45 (1) &6.45 (1) &6.45 (1)  &6.5  (1)\\
${10^{6}}$ &${90^{\circ}}$ &  &6.2  (1) &6.3  (1) &6.35 (2) &6.35 (2) &6.3  (1) &6.3  (2)  &6.3  (1)\\
          &${100^{\circ}}$&  &6    (2) &6    (1) &6    (1) &6.05 (1) &6    (2) &6.05 (1)  &6    (1)\\
          &${120^{\circ}}$&  &5.1  (1) &5.15 (1) &5.05 (1) &5.15 (1) &5.05 (1) &5.05 (1)  &5.05 (1)\\
          &${140^{\circ}}$&  &3.4  (1) &3.45 (1) &3.5  (1) &3.5  (1) &3.5  (1) &3.5  (1)  &3.5  (1)\\
          &${160^{\circ}}$&  &2.7  (1) &2.7  (1) &2.7  (1) &2.7  (1) &2.7  (1) &2.7  (1)  &2.7  (1)\\
          &${180^{\circ}}$&  &2.8  (1) &2.8  (1) &2.8  (1) &2.8  (1) &2.8  (1) &2.8  (1)  &2.8  (1)\\
          &              &${\delta X_{max}}$=&3.75&3.8&3.8&3.75&3.75&3.75&3.8\\
\hline
\end{tabular}}
\end{center}	
\end{table}
\newpage

\section*{REFERENCES}
\noindent Doolin S., Blundell K.M., 2011.  The dynamics and stability of circumbinary orbits. MNRAS, 418, 2656-2668. \\
Eggleton P., Kiseleva L., 1995.  An  empirical condition for stability of hierarchical triple systems.  ApJ, 455, 640-645. \\
Funk B., Schwarz R., Pilat-Lohinger E.,S${\ddot{u}}$li ${\acute{A}}$., Dvorak R.,2009.  Stability of inclined
orbits of terrestrial planets in habitable zones. P ${\&}$ SS, 57, 434-440.\\ 
Funk B., Libert A.-S., S${\ddot{u}}$li ${\acute{A.}}$, Pilat-Lohinger E., 2011.  On the influence of the Kozai 
mechanism in habitable zones of extrasolar planetary systems.  A${\&}$A, 526, A98.\\
Georgakarakos N., 2008.  Stability criteria for hierarchical triple systems.  Celest. Mech. Dyn. Astron., 100, 151-168.\\
Harrington R.S., 1972.  Stability Criteria for Triple Stars.  Celest. Mech., 6, 322-327.\\
H${\acute{e}}$brard G. et al., 2008.  Misaligned spin-orbit in the XO-3 planetary system?.  A${\&}$A, 488, 763-770.\\
Kozai, Y., 1962.  Secular perturbations of asteroids with high inclination and eccentricity.  AJ 67, 591-598.\\
Li J., Fu Y.N., Sun Y.S., 2010.  The Hill stability of low mass binaries in hierarchical triple systems.  Celest.  Mech. Dyn. Astron., 107, 21-34.\\
McArthur B.E.,Benedict G.F., Barnes R., Martioli E., Korzennik S., Nelan E.,
Butler R.P.,2010.  New observational constraints on the υ Andromedae system with data from the Hubble Space Telescope 
and Hobby-Eberly Telescope.  ApJ, 715, 1203-1220.\\
Marzari F., Th${\acute{e}}$bault P., Scholl H., 2009.  Planet formation in highly inclined binaries.  
A${\&}$A, 507, 505-511.\\
Mikkola S., 1997.  Practical symplectic methods with time transformation for the few-body problem. 
 Celest. Mech. Dyn. Astron., 67, 145-165.\\
Pilat-Lohinger E., 2010.  On the stability of planets in the habitable zone of inclined multi-planet systems, in 
Go${\acute{z}}$dziewski K., Niedzielski  A., Schneider J. (Eds.), Extrasolar Planets in Multi-Body Systems: 
Theory and Observations.  EAS Publication Series, Cambridge University Press, 42, pp. 403-410.\\
Pilat-Lohinger E., Funk B., Dvorak R., 2003.   Stability limits in double stars. A study of inclined planetary orbits.  
A${\&}$A, 400,1085-1094.\\
Triaud A.H.M.J. et al., 2010.  Spin-orbit angle measurements for six southern transiting planets. 
New insights into the dynamical origins of hot Jupiters.  A${\&}$A, 524, A25.\\
Valtonen M., Myll${\ddot{a}}$ri A., Orlov V., Rubinov, A., 2008.  The problem of three stars: stability limit, in: Vesperini E., 
Giersz M., Sills A. (Eds), Proc. IAU Symp. 246, Dynamical evolution of dense stellar systems, Cambridge University Press,
pp. 209-217.\\
Veras D., Armitage P.J., 2004.  The dynamics of two massive planets on inclined orbits.  
Icarus, 172, 349-371.\\
Winn J.N. et al. 2010.  The Oblique orbit of the super-Neptune HAT-P-11b.  ApJL, 723, L223-L227.\\
Winn J.N., Johnson J.A., Albrecht S., Howard A.W., Marcy G.W., Crossfield I.J.,
Holman M.J., 2009.  HAT-P-7: A Retrograde or Polar Orbit, and a Third Body.  ApJL, 703, L99-L103.\\

\end{document}